\documentclass[journal,a4paper]{IEEEtran}
\usepackage[T1]{fontenc}
\usepackage{layouts}

\usepackage{cite}

\usepackage[pdftex,colorlinks=true,bookmarks=true,citecolor=blue,urlcolor=blue]{hyperref} 

\ifCLASSINFOpdf
  \usepackage[pdftex]{graphicx}
   \graphicspath{{./figs/}}
  \DeclareGraphicsExtensions{.pdf,.jpeg,.png}
\else
\fi
\usepackage{amsmath}
\usepackage{amsthm}
\usepackage[cmintegrals]{newtxmath}
\usepackage{bm}

\usepackage{algorithmic}

\usepackage{array}

\ifCLASSOPTIONcompsoc
  \usepackage[caption=false,font=normalsize,labelfont=sf,textfont=sf]{subfig}
\else
  \usepackage[caption=false,font=footnotesize]{subfig}
\fi
\usepackage{mathrsfs}  
\usepackage{amsbsy}

\newcommand{\field}[1]{\mathbb{#1}}
\newcommand{\fs}[1]{\mathsf{#1}}

\DeclareMathOperator{\diag}{diag}

\newcommand{\tp}{\intercal}

\newcommand{\ovl}[1]{\overline{#1}}
\newcommand{\bigO}[1]{\mathop{O}(#1)}

\let\Re\relax
\DeclareMathOperator{\Re}{Re}
\let\Im\relax
\DeclareMathOperator{\Im}{Im}
\newcommand{\vv}[1]{\mathbf{#1}}
\newcommand{\vs}[1]{\boldsymbol{#1}}

\DeclareMathOperator{\sech}{sech}

\newcommand{\wtilde}[1]{\widetilde{#1}}

\usepackage{enumitem}

\begin{document}
%
\title{Higher Order Convergent Fast Nonlinear Fourier Transform}
%
%
%


\author{Vishal Vaibhav
\thanks{Email:~\tt{vishal.vaibhav@gmail.com}}
}

\IEEEpubid{\begin{minipage}[t]{\textwidth}
\vskip3em
\tiny
\centering
\copyright~2017 IEEE. Personal use of this material is permitted. Permission 
from IEEE must be obtained for all other uses, including reprinting/republishing this 
material for advertising or promotional purposes, collecting new collected works for 
resale or redistribution to servers or lists, or reuse of any copyrighted component 
of this work in other works.
\end{minipage}}


\maketitle

\begin{abstract}
It is demonstrated is this letter that linear multistep methods for 
integrating ordinary differential equations can be used to develop a family of 
fast forward scattering algorithms with higher orders of convergence. Excluding the cost of 
computing the discrete eigenvalues, the 
nonlinear Fourier transform (NFT) algorithm thus obtained has a complexity 
of $\bigO{KN+C_pN\log^2N}$ such that the error vanishes as $\bigO{N^{-p}}$ where 
$p\in\{1,2,3,4\}$ and $K$ is the number of eigenvalues. Such 
an algorithm can be potentially useful for the recently proposed 
NFT based modulation methodology for optical fiber communication. The exposition
considers the particular case of the backward differentiation formula ($C_p=p^3$) 
and the implicit Adams method ($C_p=(p-1)^3$) of which the latter proves to be
the most accurate family of methods for fast NFT.
\end{abstract}


%
\IEEEpeerreviewmaketitle

\section{Introduction}
This paper deals with the algorithmic aspects of the nonlinear Fourier
transform (NFT) based modulation scheme which aims at exploiting the nonlinear
Fourier spectrum (NF) for optical fiber
communication~\cite{Yousefi2014compact}. These novel
modulation~\cite{TPLWFK2017,FGT2017} techniques can be viewed as an extension of the original ideas of 
Hasegawa and Nyu who proposed what they coined as \emph{eigenvalue communication} in the early 
1990s~\cite{HN1993}. One of the key ingredients in various NFT-based modulation 
techniques is the fast forward NFT which can be used to decode information
encoded in the discrete and/or 
the continuous part of the nonlinear Fourier spectrum. A thorough description of the discrete
framework (based on one-step methods) for various fast forward/inverse NFT algorithms was presented 
in~\cite{V2017INFT1} where it
was shown that one can achieve a complexity of $\bigO{N\log^2N}$ in
computing the scattering coefficients in the discrete form. If the eigenvalues
are known beforehand, then the NFT has an overall complexity of
$\bigO{KN+N\log^2N}$ such that the error vanishes as $\bigO{N^{-2}}$
where $N$ is the number of samples of the signal and $K$ is the number of
eigenvalues. Interestingly enough, the complexity of the fast inverse NFT
proposed in~\cite{VW2017OFC,V2017JLTmain} also turns out to be $\bigO{KN+N\log^2N}$ with
error vanishing as $\bigO{N^{-2}}$.

In this letter, we present new fast forward scattering algorithms where the
complexity of computing the discrete scattering coefficients is 
$\bigO{C_p N\log^2N}$. If the eigenvalues are known beforehand, the NFT of a
given signal can be computed with a complexity of $\bigO{KN+C_pN\log^2N}$ such 
that the error vanishes as $\bigO{N^{-p}}$ where 
($p\in\{1,2,3,4\}$) and $K$ is the number of eigenvalues. In particular, we demonstrate 
in this work that using $m$-step ($m\in\{1,2,3,4\}$) \emph{backward
differentiation formula} (BDF) and $m$-step ($m\in\{1,2,3\}$) \emph{implicit
Adams} (IA) method~\cite{HNW1993} one can obtain fast forward NFT 
algorithms with order of convergence given by $p=m$ and $p=m+1$, respectively.

The starting point of our discussion is the Zakharov and Shabat
(ZS)~\cite{ZS1972} scattering problem which can be stated as: For
$\zeta\in\field{R}$ and $\vv{v}=(v_1,v_2)^{\tp}$,
\begin{equation}\label{eq:ZS-prob}
    \vv{v}_t = -i\zeta\sigma_3\vv{v}+U(t,x)\vv{v},
\end{equation}
where $\sigma_3=\text{diag}(1,-1)$ and the potential $U(t,x)$ is defined by
$U_{11}=U_{22}=0,\,U_{12}=q(t,x)$ and $U_{21}=r(t,x)$ with $r=\kappa q^*$
($\kappa\in\{+1, -1\}$). The parameter 
$\zeta\in\field{R}$ is known as the \emph{spectral parameter}
and $q(t,x)$ is the complex-valued function associated with the slow varying envelop of
the optical field which evolves along the fiber according to the 
nonlinear Schr\"odinger equation (NSE), stated in its normalized form,
\begin{equation}\label{eq:nse}
iq_x=q_{tt}-2 \kappa |q|^2q.
\end{equation}
The NSE provides a satisfactory description of pulse propagation in an 
optical fiber in the path-averaged formulation~\cite{Agrawal2013} under low-noise conditions 
where $t$ is the retarded time and $x$ is the distance along the fiber. In the following, the 
dependence on $x$ is 
suppressed for the sake of brevity. 
Here, $q(t)$ is identified as the \emph{scattering potential}. The 
solution of the ZS scattering 
problem~\eqref{eq:ZS-prob} consists in finding the so called 
\emph{scattering coefficients} which are defined through 
special solutions of~\eqref{eq:ZS-prob} known as the \emph{Jost solutions}.
The Jost solutions of the \emph{first kind}, denoted
by $\vs{\psi}(t;\zeta)$, has the asymptotic behavior 
$\vs{\psi}(t;\zeta)e^{-i\zeta t}\rightarrow(0,1)^{\tp}$ as $t\rightarrow\infty$. 
The Jost solutions of the \emph{second kind}, denoted by $\vs{\phi}(t,\zeta)$, has 
the asymptotic behavior $\vs{\phi}(t;\zeta)e^{i\zeta t}\rightarrow(1,0)^{\tp}$ 
as $t\rightarrow-\infty$.

For the focusing NSE (i.e., $\kappa=-1$ in~\eqref{eq:nse}), the 
nonlinear Fourier spectrum for the potential $q(t)$ comprises a \emph{discrete} and 
a \emph{continuous spectrum}. The discrete spectrum consists 
of the so-called \emph{eigenvalues} $\zeta_k\in\field{C}_+$, such that 
$a(\zeta_k)=0$, and, the \emph{norming constants} $b_k$ such that 
$\vs{\phi}(t;\zeta_k)=b_k\vs{\psi}(t;\zeta_k)$. Note that $(\zeta_k,\,b_k)$
describes a \emph{bound state} or a \emph{solitonic state}
associated with the potential. For convenience, let the
discrete spectrum be denoted by the set
\begin{equation}
\mathfrak{S}_K=\{(\zeta_k,\,b_k)\in\field{C}^2|\,\Im{\zeta_k}>0,\,k=1,2,\ldots,K\}.
\end{equation}
Note that for the defocussing NSE (i.e., $\kappa=+1$ in~\eqref{eq:nse}), the discrete spectrum 
is empty. The continuous spectrum, also referred to as the \emph{reflection coefficient}, is 
defined by $\rho(\xi)={b(\xi)}/{a(\xi)}$ for $\xi\in\field{R}$.

\begin{figure*}[th!]
\centering
\includegraphics[width=1\textwidth]{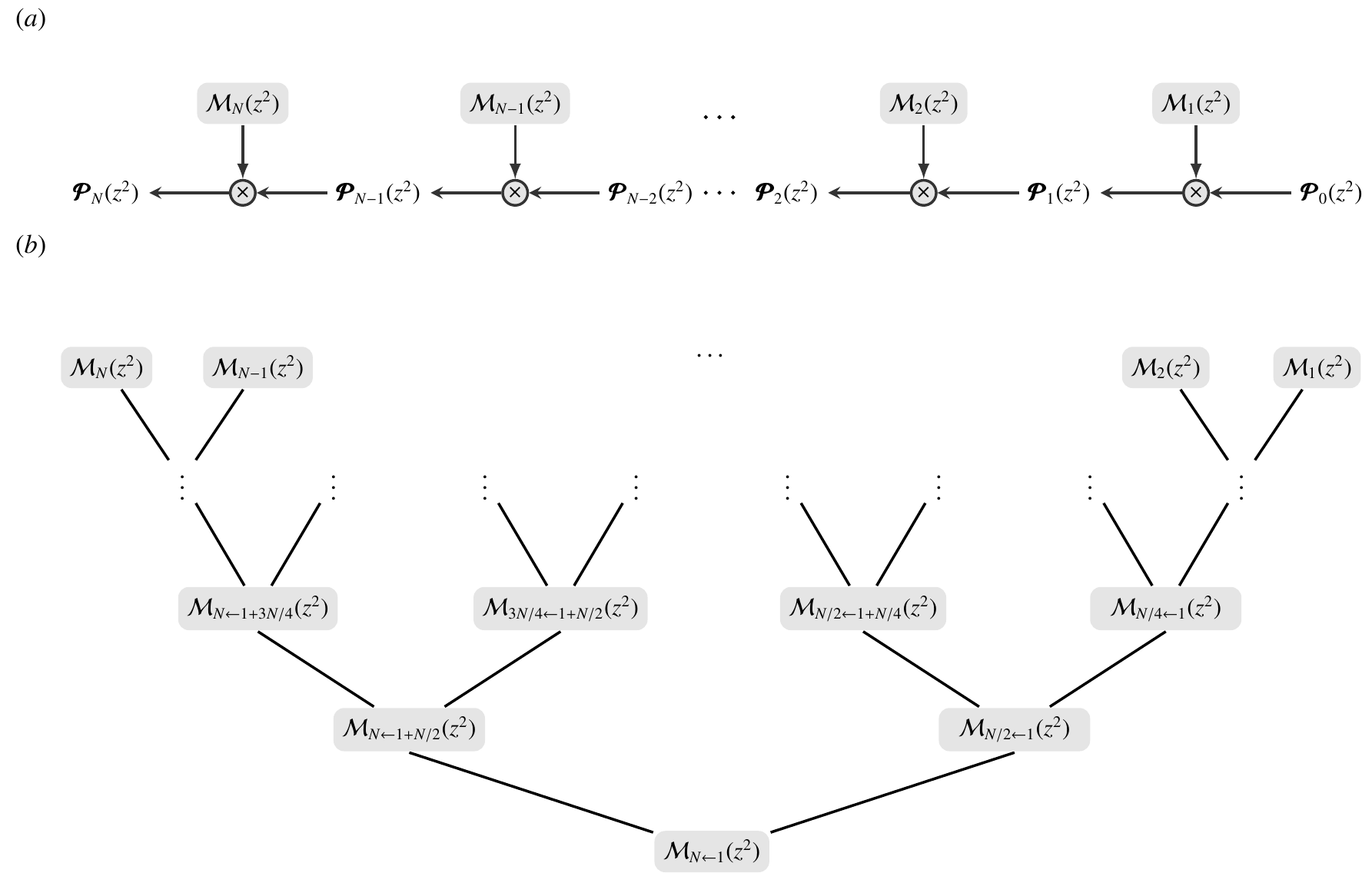}
\caption{\label{fig:schema} The figure shows the sequential approach to forward
scattering in $(a)$. A more efficient
approach is the divide-and-conquer strategy shown in $(b)$ where the transfer matrices
$\{\mathcal{M}_n(z^2)\}$ are multiplied pairwise culminating in the full transfer
matrix $\mathcal{M}_{N\leftarrow 1}(z^2)$. All polynomial products are formed using the FFT algorithm. Here
$\mathcal{M}_{n\leftarrow m}(z^2)$ denotes the cumulative 
transfer matrix
$\mathcal{M}_{n}(z^2)\times\ldots\times\mathcal{M}_{m+1}(z^2)\times\mathcal{M}_{m}(z^2)$.}
\end{figure*}
The letter first discusses the numerical discretization based on linear
multistep methods, BDF and IA, along with the algorithmic aspects. This is
followed by numerical experiments that verify the expected behavior of the 
algorithms.

\begin{figure*}[th!]
\centering
\includegraphics[width=1\textwidth]{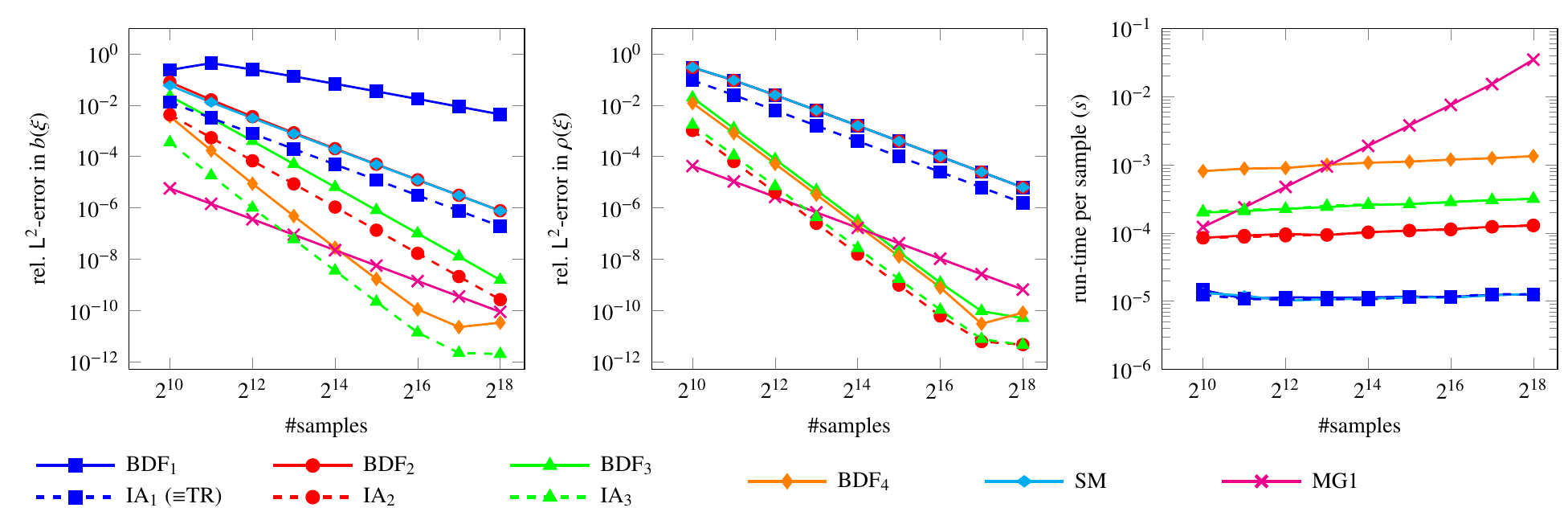}
\caption{\label{fig:results} The figure shows a comparison of convergence
behavior and run-time of NFT algorithms based on the discretization schemes, namely, BDF$_m$
($m\in\{1,2,3,4\}$), IA$_m$ ($m\in\{1,2,3\}$), SM and MG1 (see
Sec.~\ref{sec:num-exp}). The method IA$_1$ is identical to the trapezoidal rule (TR).}
\end{figure*}
The letter first discusses the numerical discretization based on linear
multistep methods, BDF and IA, along with the algorithmic aspects. This is
followed by numerical experiments that verify the expected behavior of the 
algorithms.

\section{The Numerical Scheme}
In order to develop the numerical scheme, we begin with the transformation 
$\tilde{\vv{v}}=e^{i\sigma_3\zeta t}\vv{v}$
so that~\eqref{eq:ZS-prob} becomes
\begin{equation}\label{eq:exp-int}
\tilde{\vv{v}}_t=\wtilde{U}\tilde{\vv{v}},\quad
\wtilde{U}=e^{i\sigma_3\zeta t}Ue^{-i\sigma_3\zeta t}
=\begin{pmatrix}
0 & qe^{2i\zeta t}\\
re^{-2i\zeta t} & 0
\end{pmatrix}.
\end{equation}
In order to discuss the discretization scheme, we take an equispaced grid defined 
by $t_n= T_1 + nh,\,\,n=0,1,\ldots,N,$ with $t_{N}=T_2$ where $h$ is the grid spacing.
Define $\ell_-,\ell_+\in\field{R}$ such that $h\ell_-= -T_1$, $h\ell_+= T_2$.
Further, let us define $z=e^{i\zeta h}$. For the potential functions sampled on the 
grid, we set $q_n=q(t_n)$, $r_{n}=r(t_n)$, $U_{n}=U(t_n)$ and 
$\wtilde{U}_{n}=\wtilde{U}(t_n)$. Discretization using the $m$-step BDF scheme
($m\in\{1,2,3,4\}$) reads as 
\begin{equation}
\sum_{s=0}^m\alpha_s\wtilde{\vv{v}}_{n+s} =
h\beta\wtilde{U}_{n+m}\tilde{\vv{v}}_{n+m}
\end{equation}
where $\vs{\alpha} = (\alpha_0,\alpha_1,\ldots,\alpha_m)$ and $\beta$ are known
constants~\cite[Chap.~III.1]{HNW1993}. Discretization using the
$m$-step IA method ($m\in\{1,2,3\}$) reads as
\begin{equation}
\tilde{\vv{v}}_{n+m}-\tilde{\vv{v}}_{n+m-1}=
h\sum_{s=0}^m\beta_s\wtilde{U}_{n+s}\tilde{\vv{v}}_{n+s}
\end{equation}
where $\vs{\beta} = (\beta_0,\beta_1,\ldots,\beta_m)$ are known constants~\cite[Chap.~III.1]{HNW1993}.
Both of these methods lead to a transfer matrix
$\mathcal{M}_{n+m}(z^2)\in\field{C}^{2m\times 2m}$ of the form
\begin{multline}
{\mathcal{M}}_{n+m}(z^2) = \\
\begin{pmatrix}
\gamma_{m-1}M^{(1)}_{n+m}&\gamma_{m-2}M^{(2)}_{n+m}&
\ldots&\gamma_{1}M^{(m-1)}_{n+m}&\gamma_{0}M^{(m)}_{n+m}\\
\sigma_0 &0&\ldots&0&0\\
0&\sigma_0&\ldots&0&0\\
\vdots&\vdots &\ddots & \vdots& \vdots\\
0&0&\ldots&\sigma_0&0
\end{pmatrix},
\end{multline}
where $\sigma_0=\diag(1,1)$ and $M^{(s)}_{n+m}=M^{(s)}_{n+m}(z^2)\in\field{C}^{2\times2}$ so that
\begin{equation}
\pmb{\mathcal{W}}_{n+m}={\mathcal{M}}_{n+m}(z^2)\pmb{\mathcal{W}}_{n+m-1}
\end{equation}
where $\vv{w}_n=z^{n}\vv{v}_n$ and 
$\pmb{\mathcal{W}}_{n}=(\vv{w}_{n},\vv{w}_{n-1},\ldots,\vv{w}_{n-m+1})^{\tp}\in\field{C}^{2m}$. 
For BDF schemes, we may set $\alpha_m\equiv1$. Further, setting 
$Q_{n}=(h\beta)q_{n}$, $R_{n}=(h\beta)r_{n}$ and $\Theta_n =
1-Q_{n}R_{n}$, we have $\gamma_s=-\alpha_s$ together with
\begin{equation}
M^{(s)}_{n+m}(z^2)=\frac{1}{\Theta_{n+m}}\begin{pmatrix}
1&z^{2s}Q_{n+m}\\
R_{n+m} & z^{2s}
\end{pmatrix}.
\end{equation}
For the IA methods, we have
\begin{multline}
M^{(1)}_{n+m}(z^2)=\Theta^{-1}_{n+m}\times\\
\begin{pmatrix}
1+z^{2}\bar{\beta}_{m-1}R_{n+m-1}Q_{n+m} &z^{2}Q_{n+m} + \bar{\beta}_{m-1}Q_{n+m-1}\\
R_{n+m}+z^{2}\bar{\beta}_{m-1}R_{n+m-1} &
z^{2}+\bar{\beta}_{m-1}R_{n+m}Q_{n+m-1}
\end{pmatrix},
\end{multline}
where $Q_{n}=(h\beta_m)q_{n}$, $R_{n}=(h\beta_m)r_{n}$, $\Theta_n =
1-Q_{n}R_{n}$.
Also,
\begin{equation}
M^{(m-s)}_{n+m}(z^2)=\frac{1}{\Theta_{n+m}}\begin{pmatrix}
    z^{2(m-s)}R_{n+s}Q_{n+m} &Q_{n+s}\\
    z^{2(m-s)}R_{n+s} & R_{n+m}Q_{n+s}
\end{pmatrix},
\end{equation}
with $\gamma_{m-1}=1$ and $\gamma_{s}=\ovl{\beta}_s$ for
$s=0,1,\ldots,m-2$ where
\begin{equation}
\ovl{\vs{\beta}} = \vs{\beta}/\beta_m=(\ovl{\beta}_0,\ovl{\beta}_1,\ldots,1). 
\end{equation}

Let us consider the Jost solution $\vs{\phi}(t;\zeta)$. We assume that $q_n=0$ for 
$n=-m+1, -m+2,\ldots,0$ so that 
$\vs{\phi}_n=z^{\ell_-}z^{-n}(1,0)^{\tp}$ for $n=-m+1,-m+2,\ldots,0$. In order 
to express the discrete approximation to the Jost solutions, let us define 
the vector-valued polynomial
\begin{equation}\label{eq:poly-vec}
\vv{P}_n(z^2)
=\begin{pmatrix}
P^{(n)}_{1}(z^2)\\
P^{(n)}_{2}(z^2)
\end{pmatrix}\\
=\sum_{j=0}^{n}
\vv{P}^{(n)}_{j}z^{2j}
=\sum_{j=0}^{n}
\begin{pmatrix}
P^{(n)}_{1,j}\\
P^{(n)}_{2,j}
\end{pmatrix}^{\tp}z^{2j},
\end{equation}
such that $\vs{\phi}_n = z^{\ell_-}z^{-n}\vv{P}_n(z^2)$. 
The initial condition
works out to be 
\begin{equation}
\pmb{\mathcal{W}}_{0}=z^{\ell_-}
\begin{pmatrix}
\vs{\phi}_{0}\\
z\vs{\phi}_{-1}\\
\vdots\\
z^{-m+1}\vs{\phi}_{-m+1}
\end{pmatrix}
=z^{\ell_-}
\begin{pmatrix}
\vs{P}_{0}(z^2)\\
\vs{P}_{-1}(z^2)\\
\vdots\\
\vs{P}_{-m+1}(z^2)
\end{pmatrix}
\in\field{C}^{2m},
\end{equation}
yielding the recurrence relation
\begin{equation}
\pmb{\mathcal{P}}_{n+m}(z^2)={\mathcal{M}}_{n+m}(z^2)\pmb{\mathcal{P}}_{n+m-1}(z^2),
\end{equation}
where 
$\pmb{\mathcal{P}}_{n}(z^2) =
(\vv{P}_{n}(z^2),\vv{P}_{n-1}(z^2),\ldots,\vv{P}_{n-m+1}(z^2))^{\tp}\in\field{C}^{2m}$.
The discrete approximation to the scattering coefficients is obtained from the scattered
field: $\vs{\phi}_{N}=(a_{N} z^{-\ell_+},b_{N} z^{\ell_+})^{\tp}$ yields
$a_{N}(z^2)={P}^{(N)}_1(z^2)$ and
$b_{N}(z^2)=(z^2)^{-\ell_{+}}{P}^{(N)}_2(z^2)$. The quantities $a_{N}$ and $b_{N}$ are referred to as the 
\emph{discrete scattering coefficients} uniquely defined for $\Re\zeta\in
[-{\pi}/{2h},\,{\pi}/{2h}]$.

Finally, let us mention that, for $\zeta$ varying over a compact domain, the
error in the computation of the scattering coefficients can be shown to be
$\bigO{N^{-p}}$ provided that $q(t)$ is at least $p$-times differentiable~\cite[Chap.~III]{HNW1993}.

\subsection{Fast Forward Scattering Algorithm}
It is evident from the preceding paragraph that
the forward scattering step requires forming the cumulative product:
${\mathcal{M}}_{N}(z^2)\times{\mathcal{M}}_{N-1}(z^2)\times\ldots
\times{\mathcal{M}}_{2}(z^2)\times{\mathcal{M}}_{1}(z^2)$. Let $\bar{m}$ 
denote the nearest base-$2$ number greater than or equal
to $(m +1)$, then pairwise multiplication using FFT~\cite{Henrici1979} yields the 
recurrence relation for the complexity $\varpi(n)$ of computing the scattering
coefficients with $n$ samples: 
$\varpi(n) = 8m^3\nu(\bar{m}n/2)+2\varpi(n/2),\,\,n=2,\,4,\,\ldots,\,N,$ 
where $\nu(n)=\bigO{n\log n}$ is the cost of multiplying two polynomials of
degree $n-1$ (ignoring the cost of additions). Solving the
recurrence relation yields $\varpi(N)=\bigO{m^3N\log^2N}$.

\subsubsection{Computation of the continuous spectrum}
The computation of the continuous spectrum requires evaluation the polynomial
$b_N(z^2)$ and $a_N(z^2)$ on the unit circle $|z|=1$, say, at $N$ points. This can be done
efficiently using the FFT algorithm with complexity $\bigO{N\log N}$. Therefore,
the overall complexity of computation of the continuous spectrum easily works to
be $\bigO{m^3N\log^2N}$.
\subsubsection{Computation of the norming constants}
Let us assume that the discrete eigenvalues are known by design\footnote{Given
that the best polynomial root-finding algorithms still require $\bigO{N^2}$
operations, we would at this stage favor a system design which avoids 
having to compute eigenvalues.}. Therefore the
only part of the discrete spectrum still to be computed are the norming
constants. A method of computing the norming constants corresponding to
arbitrary eigenvalues is presented in~\cite{V2017INFT1} which has an additional complexity 
of $\bigO{KN}$ where $K$ is the number of eigenvalues. This method can be
employed here as well because it uses no information regarding how the discrete
scattering coefficients were computed.

\begin{figure}[!th]
\centering
\includegraphics[scale=1]{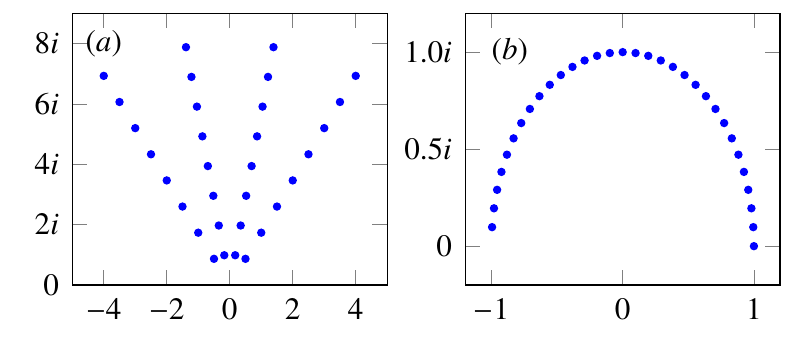}%
\caption{The figure depicts $\mathfrak{S}_{32}$ defined by~\eqref{eq:ds-test-ms}, 
where the eigenvalues and the norming constants
are shown in (a) and (b), respectively.\label{fig:ds-test}}
\end{figure}
\section{Numerical Experiments: Test for Convergence and Complexity}
\label{sec:num-exp}
\subsection{Secant-hyperbolic potential}
A test for verifying the order of convergence and complexity can be readily
designed using the well-known secant-hyperbolic potential given by 
$q(t) = A \sech{t}$, ($\kappa=-1$). The scattering coefficients are given by~\cite{SY1974} 
\begin{equation}
\begin{split}
&a(\xi)=\frac{\left[\Gamma\left(0.5-i\xi\right)\right]^2}
{\Gamma\left(A+0.5-i\xi\right)\Gamma\left(-A+0.5-i\xi\right)},\\
&b(\xi)= -\sin\pi A\sech\pi\xi,
\end{split}
\end{equation}
so that the reflection coefficient is given by $\rho(\xi)=b(\xi)/a(\xi)$.
We set $A=4.4$. Let $\Omega_h=[-{\pi}/{2h},\,{\pi}/{2h}]$; then, the error in computing 
$b(\xi)$ is quantified by  
\begin{equation}
e_{\text{rel.}}=\|b(\xi)-b_N(\xi)\|_{\fs{L}^2(\Omega_h)}/\|b(\xi)\|_{\fs{L}^2(\Omega_h)},
\end{equation}
where the integrals are computed using the trapezoidal rule. Similar consideration applies 
to $\rho(\xi)$. For the purpose of benchmarking, we use the Split-Magnus (SM)
and Magnus method with one-point Gauss quadrature (MG1) discussed in~\cite[Sec.~IV]{V2017INFT1}).
Note that the complexity of SM is $\bigO{N\log^2N}$ in computing
the scattering coefficients while that of MG1 is $\bigO{N^2}$. The order of
convergence for SM and MG1 both is $\bigO{N^{-2}}$. The numerical results
are plotted in Fig.~\ref{fig:results} where it is evident that $m$-step BDF
(labeled BDF$_m$) as well as the $m$-step IA (labeled IA$_m$ where IA$_1$ is
identical to trapezoidal rule (TR)) schemes have better 
convergence rates with increasing $m$. The improved accuracy, however, comes at a 
price of increased complexity which is evidently not so prohibitive
(besides, room for improvements in the implementation does exist). The IA
methods are clearly superior to that of BDF in terms of accuracy while keeping the
complexity same.  

\begin{figure*}[ht!]
\centering
\includegraphics[width=1\textwidth]{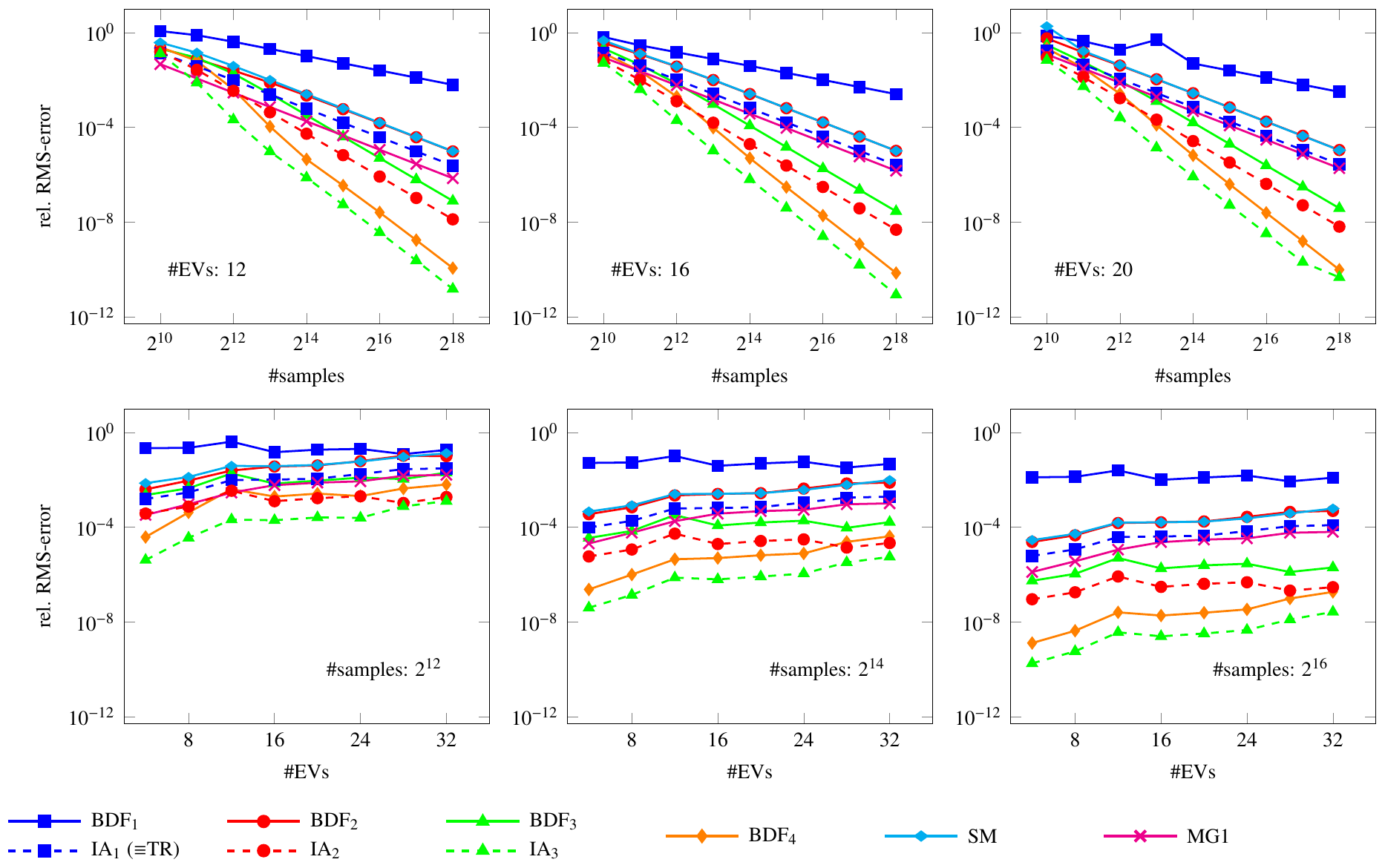}
\caption{\label{fig:results-nconst} The figure shows a comparison of convergence
behavior of NFT algorithms for computing the norming constant 
based on the discretization schemes, namely, BDF$_m$
($m\in\{1,2,3,4\}$), IA$_m$ ($m\in\{1,2,3\}$), SM and MG1 (see
Sec.~\ref{sec:num-exp}). The method IA$_1$ is identical to the trapezoidal rule (TR).}
\end{figure*}
\subsection{Multisolitons}
Arbitrary multisoliton solutions can be computed using the classical
Darboux transformation (CDT), which allows us to test our algorithms for 
computing the norming constants. To this end, we define
an arbitrary discrete spectrum and compute the corresponding multisoliton 
solution which serves as an input to the NFT algorithms. Let $b^{(\text{num.})}_k$ 
be the numerically computed approximation to $b_k$ which corresponds to 
the eigenvalue $\zeta_k$ which we assume to be known. The error 
in the norming constants can then be quantified by
\begin{equation}
e_{\text{rel}} 
=\sqrt{\left({\sum_{k=1}^K|b^{(\text{num.})}_k-b_k|^2}\right)\biggl/{\sum_{k=1}^K|b_k|^2}}.
\end{equation}
For the discrete spectrum, the example chosen here is taken from~\cite{V2017INFT1}
which can be described as follows: Define a sequence of angles for $J\in\field{Z}_+$ by choosing
$\Delta\theta=(\pi-2\theta_0)/(J-1),\,\theta_0>0$, and 
$\theta_j=\theta_0 + (j-1)\Delta\theta,\,j=1,2,\ldots,J$ 
so that $\theta_j\in[\theta_0,\pi-\theta_0]$. Then the eigenvalues are chosen as
\begin{equation}
    \zeta_{j+J(l-1)}=le^{i\theta_j},\,l=1,2,\ldots,8,\,j=1,2,\ldots,J.
\end{equation}
Further, the norming constants are chosen as
\begin{equation}
    b_j=e^{i\pi (j-1)/(8J-1)},\,j=1,2,\ldots,8J.    
\end{equation}
For this test, we set $\theta_0=\pi/3$ and $J=4$. Then we consider a sequence
of discrete spectra defined as
\begin{equation}
\label{eq:ds-test-ms}
\mathfrak{S}_{K} = \{(\zeta_k, b_k),\,k=1,2,\ldots,K\},
\end{equation}
where $K=4,8,\ldots,32$ (see Fig.~\ref{fig:ds-test}). For fixed $K$,
the eigenvalues are scaled by the scaling parameter $\kappa =
2(\sum_{k=1}^K\Im\zeta_k)^{1/2}$. Let
$\eta_{\text{min}}=\min_{\{\zeta_k\}}\Im\zeta$, then the computational domain for this
example is chosen as $[-T,\,T]$ where $T={22\kappa}/\eta_{\text{min}}$. The numerical results
are plotted in Fig.~\ref{fig:ds-test} where it is evident that BDF$_m$ as well as 
IA$_m$ schemes have better 
convergence rates with increasing $m$. The IA
methods are clearly superior to that of BDF in terms of accuracy.

\section{Conclusion}
In this letter we presented a family of fast NFT algorithms based on exponential linear multistep
methods which were demonstrated to exhibit higher-order of convergence. Excluding the cost of 
computing the discrete eigenvalues, the proposed algorithms have a complexity 
of $\bigO{KN+C_pN\log^2N}$ such that the error vanishes as $\bigO{N^{-p}}$ where 
$p\in\{1,2,3,4\}$ and $K$ is the number of eigenvalues. The form of $C_p$
depends on the underlying linear multistep method.

The future research in this direction will focus on developing compatible 
fast layer-peeling schemes for the discrete systems proposed in this letter so that
higher-order convergent fast inverse NFT algorithms could be developed.

\bibliographystyle{IEEEtran}

\begin{thebibliography}{10}
\providecommand{\url}[1]{#1}
\csname url@samestyle\endcsname
\providecommand{\newblock}{\relax}
\providecommand{\bibinfo}[2]{#2}
\providecommand{\BIBentrySTDinterwordspacing}{\spaceskip=0pt\relax}
\providecommand{\BIBentryALTinterwordstretchfactor}{4}
\providecommand{\BIBentryALTinterwordspacing}{\spaceskip=\fontdimen2\font plus
\BIBentryALTinterwordstretchfactor\fontdimen3\font minus
  \fontdimen4\font\relax}
\providecommand{\BIBforeignlanguage}[2]{{%
\expandafter\ifx\csname l@#1\endcsname\relax
\typeout{** WARNING: IEEEtran.bst: No hyphenation pattern has been}%
\typeout{** loaded for the language `#1'. Using the pattern for}%
\typeout{** the default language instead.}%
\else
\language=\csname l@#1\endcsname
\fi
#2}}
\providecommand{\BIBdecl}{\relax}
\BIBdecl

\bibitem{Yousefi2014compact}
M.~I. Yousefi and F.~R. Kschischang, ``Information transmission using the
  nonlinear {F}ourier transform, {P}art {I},'' \emph{IEEE Trans. Inf. Theory},
  vol.~60, no.~7, pp. 4312--4369, 2014.

\bibitem{TPLWFK2017}
S.~K. Turitsyn, J.~E. Prilepsky, S.~T. Le, S.~Wahls, L.~L. Frumin, M.~Kamalian,
  and S.~A. Derevyanko, ``Nonlinear {F}ourier transform for optical data
  processing and transmission: advances and perspectives,'' \emph{Optica},
  vol.~4, no.~3, pp. 307--322, Mar 2017.

\bibitem{FGT2017}
L.~L. Frumin, A.~A. Gelash, and S.~K. Turitsyn, ``New approaches to coding
  information using inverse scattering transform,'' \emph{Phys. Rev. Lett.},
  vol. 118, p. 223901, 2017.

\bibitem{HN1993}
A.~Hasegawa and T.~Nyu, ``Eigenvalue communication,'' \emph{J. Lightwave
  Technol.}, vol.~11, no.~3, pp. 395--399, Mar 1993.

\bibitem{V2017INFT1}
V.~Vaibhav, ``Fast inverse nonlinear {F}ourier transformation using exponential
  one-step methods: {D}arboux transformation,'' \emph{Phys. Rev. E}, vol.~96,
  p. 063302, 2017.

\bibitem{VW2017OFC}
V.~Vaibhav and S.~Wahls, ``Introducing the fast inverse {NFT},'' in
  \emph{Optical Fiber Communication Conference}.\hskip 1em plus 0.5em minus
  0.4em\relax Los Angeles, CA, USA: Optical Society of America, 2017, p.
  Tu3D.2.

\bibitem{V2017JLTmain}
V.~Vaibhav, ``Fast inverse nonlinear {F}ourier transformation,'' 2017,
  {arXiv}:1706.04069[math.NA].

\bibitem{HNW1993}
E.~Hairer, S.~P. N{\o}rsett, and G.~Wanner, \emph{Solving Ordinary Differential
  Equations I: Nonstiff Problems}, ser. Springer Series in Computational
  Mathematics.\hskip 1em plus 0.5em minus 0.4em\relax Berlin: Springer, 1993.

\bibitem{ZS1972}
V.~E. Zakharov and A.~B. Shabat, ``Exact theory of two-dimensional
  self-focusing and one-dimensional self-modulation of waves in nonlinear
  media,'' \emph{Sov. Phys. JETP}, vol.~34, pp. 62--69, 1972.

\bibitem{Agrawal2013}
G.~P. Agrawal, \emph{Nonlinear Fiber Optics}, 3rd~ed., ser. Optics and
  Photonics.\hskip 1em plus 0.5em minus 0.4em\relax New York: Academic Press,
  2013.

\bibitem{Henrici1979}
P.~Henrici, ``Fast {F}ourier methods in computational complex analysis,''
  \emph{SIAM Review}, vol.~21, no.~4, pp. 481--527, 1979.

\bibitem{SY1974}
J.~Satsuma and N.~Yajima, ``B. initial value problems of one-dimensional
  self-modulation of nonlinear waves in dispersive media,'' \emph{Prog. Theor.
  Phys. Suppl.}, vol.~55, pp. 284--306, 1974.

\end{thebibliography}

\providecommand{\noopsort}[1]{}\providecommand{\singleletter}[1]{#1}%

\end{document}